\documentclass[aps,preprint,onecolumn,floatfix,superscriptaddress,nofootinbib]{revtex4-1}
\usepackage{multirow}
\usepackage{makecell}
\usepackage{amsmath}
\usepackage{graphicx,bm}
\usepackage[colorlinks,citecolor=blue]{hyperref}
\usepackage[caption=false]{subfig}
\usepackage{slashed}
\usepackage{ulem}
\usepackage{fancyhdr}
\usepackage{fancyvrb}
\usepackage[absolute,overlay]{textpos}

\begin{document}

\begin{textblock}{4}(12,0.01)   
  \includegraphics[width=1cm]{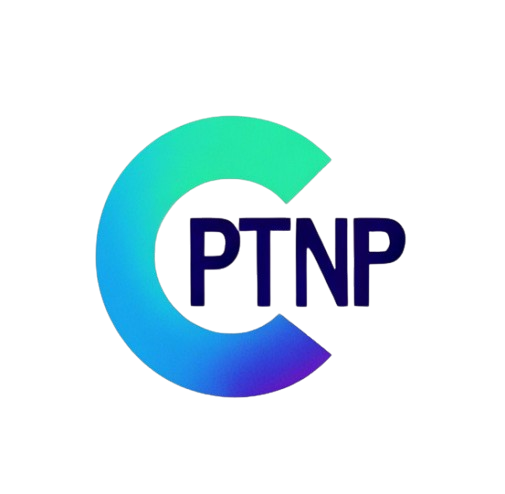}  
\end{textblock}
\begin{textblock}{6}(9.5,0.2)
  \raggedleft
  {\text{CPTNP-2025-012}}  
\end{textblock}

\title{Neutrinophilic Super-Resonant Dark Matter}

\author{Murat Abdughani}
\email{mulati@xju.edu.cn}
\affiliation{School of Physical Science and Technology, Xinjiang University, Urumqi 830046, China}

\author{Shao-Song Tang}
\email{107552300734@stu.edu.cn}
\affiliation{School of Physical Science and Technology, Xinjiang University, Urumqi 830046, China}

\author{Kadirya Tursun}
\email{kadirya@xao.ac.cn}
\affiliation{Xinjiang Astronomical Observatory, Chinese Academy of Sciences, Urumqi 830011, P. R. China}

\author{Bin Zhu}
\email{zhubin@mail.nankai.edu.cn}
\affiliation{Department of Physics, Yantai University, Yantai 264005, P. R. China}

\begin{abstract}
Dark matter (DM) annihilation can be significantly enhanced through narrow resonances or the Sommerfeld enhancement effect, with both mechanisms potentially combining in a super-resonant annihilation process. In such scenarios, the conventional assumption that kinetic equilibrium persists until chemical decoupling may not hold, leading to substantial impacts on the final DM relic density. However, a strongly enhanced annihilation cross section into Standard Model particles, except neutrinos, is constrained by cosmic microwave background observations. We thus investigate DM annihilation into neutrino pair final states, focusing on the role of kinetic decoupling. We solve the coupled Boltzmann equations to determine the relic density and constrain the parameter space using current experimental data, while also forecasting the sensitivity of future experiments.
\end{abstract}

\date{\today}
\maketitle
\newpage

\section{Introduction} \label{sec:Intro}
The relic density of dark matter (DM), quantified as $\Omega_{\mathrm{DM}} h^2 \approx 0.12$ from precision cosmological observations~\cite{Planck:2018vyg, DESI:2024mwx}, serves as a cornerstone of modern cosmology. This value supports the $\Lambda$CDM paradigm and provides a critical benchmark for testing particle physics models beyond the Standard Model (SM). Any viable DM candidate must reproduce this relic abundance through its production mechanism while remaining consistent with the cosmic microwave background (CMB) anisotropies~\cite{Planck:2018jri} and large-scale structure formation~\cite{Springel:2005nw}. Deviations from the observed value would either overclose the universe or fail to explain gravitational lensing and galactic rotation curves. Consequently, precise relic density calculations—especially in scenarios where conventional approximations break down—are essential to test the DM models.

A leading explanation for the observed DM relic density is that DM particles, as elementary particles, are thermally produced shortly after inflation~\cite{Sarkar:1995dd, Cirelli:2024ssz, Guth:1980zm, Tenkanen:2019aij}. Initially relativistic, these particles maintain thermal equilibrium with SM particles through frequent interactions~\cite{Bertone:2004pz, Poulin:2019omz}. As the universe expands, the DM annihilation rate decreases, and SM particles do not possess sufficient energy to produce DM particles, leading to a comoving DM number density that remains fixed after freeze-out~\cite{Edsjo:1997bg, Harigaya:2016nlg, Dienes:2017zjq, Jaramillo:2020dde, Marfatia:2020bcs}. If the DM mass and interaction strengths are approximately at the weak scale, it naturally yields the observed relic density, a phenomenon known as the Weakly Interacting Massive Particle (WIMP) miracle~\cite{Ellis:1983ew, Jungman:1995df}.

In most cases, it is sufficient to consider only the first moment of the DM phase-space density, namely the number density $n_{\mathrm{DM}}$, when solving the Boltzmann equation to determine the relic density, provided that kinetic equilibrium is maintained until chemical decoupling occurs. However, when kinetic and chemical decoupling occurs concurrently, it becomes inevitable to include the second moment of the phase-space density or solve the full Boltzmann equation to achieve more accurate results~\cite{Duch:2017nbe, Kamada:2017gfc, Abe:2020obo, Binder:2017rgn, 2103.01944,2301.12199,2404.12019}. For instance, in Ref.~\cite{2103.01944}, three well-known cases~\cite{Griest:1990kh} where the standard relic density calculation is inadequate are analyzed with greater precision, achieving differences of up to 1-2 orders of magnitude in the results when the case of kinetic decoupling occurs before chemical decoupling is considered.

The need for such refined calculations often arises from specific physical processes that complicate the dynamics of DM interactions. In particular, the annihilation cross section, which governs the rate of DM particle interactions, plays a pivotal role in determining the relic density and can be significantly influenced by the Sommerfeld effect~\cite{Sommerfeld:1931qaf, Feng:2010zp, Arkani-Hamed:2008hhe} or narrow resonance~\cite{Ibe:2008ye, Duch:2017nbe, Belanger:2024bro,2309.12043,2504.00383}. Notably, these two effects may occur concurrently, significantly amplifying the annihilation cross section via the super-resonant mechanism~\cite{Beneke:2022rjv}.  In such scenarios, due to the complexity of interactions between DM and other particles, the relic density derived from the traditional Boltzmann equation may differ substantially from that obtained using more accurate coupled or full Boltzmann equations.

Crucially, the velocity dependence of the annihilation cross-section—dictated by whether the process is $s$-wave or $p$-wave dominated—introduces additional complexity. While $s$-wave annihilation (thermally averaged annihilation cross section $\langle \sigma v\rangle \propto v^0 $) dominates in the non-relativistic regime relevant for freeze-out, $p$-wave processes ($\langle \sigma v\rangle \propto v^2 $) exhibit suppressed rates at low velocities, leading to distinct observational signatures. If DM annihilation is a $p$-wave process, the enhanced cross section may be insufficient to counteract the universe's expansion, rendering its impact on the relic density negligible~\cite{An:2016kie, Johnson:2019hsm}. Moreover, current annihilation cross sections at galactic centers are too small to produce detectable signals. Conversely, if the annihilation is an $s$-wave process, a significantly enhanced cross section during the CMB era could be excluded by Planck data~\cite{Planck:2018vyg} if the annihilation products are SM particles other than neutrino pairs. Therefore, we investigate the scenario where two DM particles annihilate into a neutrino pair final state, namely neutrinophilic DM.

The structure of this paper is as follows: in Sec.~\ref{sec:kinetic}, we derive the coupled Boltzmann equations (cBEs); in Sec.~\ref{sec:benchmark}, we present a benchmark super-resonant DM (SRDM) model; in Sec.~\ref{sec:constrain}, we provide current constraints and future prospects for the benchmark model; and we conclude in Sec.~\ref{sec:conclusion}.

\section{Out-of-kinetic-equilibrium} \label{sec:kinetic}

In cosmology, the focus is on the statistical behavior of particles rather than their individual dynamics~\cite{Scherrer:1985zt}. The evolution of the particle phase-space density $f(\mathbf{x}, \mathbf{p}, t)$ is governed by the Boltzmann equation:
\begin{equation}
    \frac{df}{dt} = C[f],
\end{equation}
where $C[f]$ represents the collision term. In a homogeneous and isotropic Friedmann-Robertson-Walker universe, the Boltzmann equation for the DM particle $\chi$, with phase-space density $f_\chi(t, p)$, is given by~\cite{2103.01944}:
\begin{equation}
    E \left( \partial_t - H p \partial_p \right) f_\chi = C_{\rm ann}[f_\chi] + C_{\rm el}[f_\chi], \label{eq:fullBoltz}
\end{equation}
where $H = \dot{a}/a$ is the Hubble parameter, $C_{el}$ is elastic scattering term, and $a$ is the scale factor. Near freeze-out, for non-relativistic DM where $f_\chi \ll 1$, the annihilation term on the right-hand side of Eq.~\eqref{eq:fullBoltz} reduces to~\cite{Griest:1990kh}:
\begin{equation}
    C_{\rm ann} = g_\chi E \int \frac{d^3 \tilde{p}}{(2\pi)^3} v_M \sigma_{\bar{\chi} \chi \to \bar{f} f} \left[ f_{\chi, {\rm eq}}(E) f_{\chi, {\rm eq}}(\tilde{E}) - f_\chi(E) f_\chi(\tilde{E}) \right],
\end{equation}
where $g_\chi$ is the internal degrees of freedom of DM, $\sigma$ is the annihilation cross section to the $\bar{f} f$ final state, $v_M$ is the Møller velocity, and $f_{\chi, {\rm eq}}(E)$ is the equilibrium phase-space density for non-relativistic DM.

When $m_f \ll m_\chi$, a second-order expansion of $C_{\rm el}$ in the momentum transfer yields a differential operator of the Fokker-Planck (FP) type~\cite{astro-ph/0607319, Bringmann:2006mu}. After relativistic corrections, the FP operator becomes:
\begin{equation}
    C_{\rm el} \simeq C_{\rm FP} = \frac{1}{2} E \gamma(T) \left[ T E \partial_p^2 + \left( \frac{2 T E}{p} + p + \frac{T p}{E} \right) \partial_p + 3 \right] f_\chi,
\end{equation}
where the momentum transfer rate is~\cite{1205.1914}:
\begin{equation}
    \gamma(T) = \frac{1}{3 g_\chi m_\chi T} \int \frac{d^3 k}{(2\pi)^3} g^\pm (w) [1 \mp g^\pm (w)] \int_{-4 k^2_{\rm cm}}^0 dt (-t) \frac{d\sigma}{dt} v, \label{eq:gamma}
\end{equation}
with $g^\pm (w) = 1/[e^{w/T} \pm 1]$ as the phase-space distribution of heat bath particles, $\frac{d\sigma}{dt}v  \equiv |\mathcal{M}|^2_{\chi f \leftrightarrow \chi f} / (64 \pi k w m_\chi^2)$ evaluated at $s \simeq m_\chi^2 + 2 w m_\chi + m_f^2$, and $k_{\rm cm}^2 \equiv m_\chi^2 k^2 / (m_\chi^2 + 2 w m_\chi + m_f^2)$.

Solving the full Boltzmann equation in Eq.~\eqref{eq:fullBoltz} is computationally intensive~\cite{Brummer:2019inq, Ala-Mattinen:2019mpa}. Instead, a hydrodynamical approach using low moments of the phase-space distribution and additional assumptions yields more tractable fluid equations~\cite{Binder:2021bmg, Binder:2017rgn, Kamada:2018hte, Cui:2020ppc}. The simplest method to compute the relic density involves considering only the lowest moment of $f_\chi$, the number density $n \equiv g_\chi \int \frac{d^3 p}{(2\pi)^3} f_\chi$, assuming kinetic decoupling occurs after chemical decoupling. For non-relativistic DM, $f_\chi = e^{(\mu - E)/T}$, with the chemical potential $\mu$ satisfying $\mu / T = \ln(n / n_{\rm eq})$, and the equilibrium number density ($\mu = 0$) is $n_{\rm eq}(T) = g_\chi \frac{m_\chi^2 T}{2 \pi^2} K_2(m_\chi / T)$, where $K_2$ is the modified Bessel function of the second kind. Substituting this $f_\chi$ into Eq.~\eqref{eq:fullBoltz} results in the well-known number density equation (nBE):
\begin{equation}
    \dot{n} + 3 H n = - \langle \sigma v \rangle_T (n^2 - n_{\rm eq}^2), \label{eq:nBE}
\end{equation}
where $\langle \sigma v \rangle_T$ is the thermally averaged annihilation cross section.

When kinetic and chemical decoupling are intertwined by each other, using Eq.~\eqref{eq:nBE} to calculate the relic density may introduce significant errors~\cite{Binder:2017rgn, Duch:2017nbe}. These discrepancies can be mitigated by incorporating the second moment of $f_\chi$, typically the DM velocity dispersion or temperature, defined as $T_\chi \equiv \frac{g_\chi}{3 n_\chi} \int \frac{d^3 p}{(2\pi)^3} \frac{p^2}{E} f_\chi$. Assuming $f_\chi = e^{(\mu - E)/T_\chi}$ with $\mu / T_\chi = \ln(n / n_{\rm eq}(T_\chi))$, the cBEs are~\cite{2103.01944,Binder:2017rgn}:
\begin{eqnarray}
    \frac{Y'}{Y} &=& \frac{s Y}{x \tilde{H}} \left[ \frac{Y_{\rm eq}^2}{Y^2} \langle \sigma v \rangle_T - \langle \sigma v \rangle_{T_\chi} \right], \nonumber \\
    \frac{y'}{y} &=& \frac{1}{x \tilde{H}} \langle C_{\rm el} \rangle_2 + \frac{s Y}{x \tilde{H}} \left[ \langle \sigma v \rangle_{T_\chi} - \langle \sigma v \rangle_{2, T_\chi} \right] \nonumber \\
    &+& \frac{s Y}{x \tilde{H}} \frac{Y_{\rm eq}^2}{Y^2} \left[ \frac{y_{\rm eq}}{y} \langle \sigma v \rangle_2 - \langle \sigma v \rangle_T \right] + 2 (1 - w) \frac{H}{x \tilde{H}}, \label{eq:cBE}
\end{eqnarray}
where $Y(x) \equiv n_\chi / s$, $x \equiv m_\chi / T$, $Y_{\rm eq}(x) = n_{\rm eq}(T) / s$, $y \equiv m_\chi T_\chi / s^{2/3}$, $s$ is the entropy density, and $w(T_\chi) \equiv 1 - \langle p^4 / E^3 \rangle_{T_\chi} / (6 T_\chi)$ measures deviation from the highly non-relativistic case ($w = 1$). The variant of the thermally averaged annihilation cross section is:
\begin{equation}
    \langle \sigma v \rangle_2 \equiv \frac{g_\chi^2}{T n_{\rm eq}} \int \frac{d^3 p \, d^3 \tilde{p}}{(2\pi)^6} \frac{p^2}{3 E} \sigma v_{\bar{\chi} \chi \to \bar{f} f} f_{\chi, {\rm eq}}(\mathbf{p}) f_{\chi, {\rm eq}}(\tilde{\mathbf{p}}).
\end{equation}
The elastic term and modified Hubble parameter are defined as:
\begin{equation}
    \langle C_{\rm el} \rangle_2 \equiv \frac{g_\chi}{3 n T_\chi} \int \frac{d^3 p}{(2\pi)^3} \frac{p^2}{E^2} C_{\rm el},
\end{equation}
\begin{equation}
    \tilde{H} \equiv \frac{H}{1 + \frac{T}{3 g_{\rm eff}} \frac{d g_{\rm eff}}{d T}},
\end{equation}
where $g_{\rm eff}$ is the entropy degrees of freedom of the background plasma.

\section{Benchmark Model} \label{sec:benchmark}

We propose a simple benchmark model featuring a Dirac fermion DM $\chi$, real vector mediators $A$ and $V$, and massive Dirac neutrinos $\nu$. The particles $A$ and $V$, with masses $m_A \ll m_\chi$ and $m_V \simeq 2 m_\chi$, mediate a long-range force and narrow resonance, respectively. The minimal interaction Lagrangian is:
\begin{equation}
    \mathcal{L} \supset - g_A \bar{\chi} \gamma^\mu \chi A_\mu - g_V \bar{\chi} \gamma^\mu \chi V_\mu - g_\nu \bar{\nu} \gamma^\mu \nu V_\mu, \label{eq:lagrangian}
\end{equation}
where $g_{A,V,\nu}$ are coupling constants. Dirac neutrinos couple to $V$ to ensure thermal equilibrium of the new particles in the early universe. The model does not originate from gauge symmetry but is phenomenologically viable after the electroweak symmetry breaking of higher-dimensional effective operators. Although $A$ does not directly couple to either neutrinos or $V$, this separation allows the annihilation cross section to factorize into the product of a Sommerfeld factor and a Breit–Wigner resonance factor, as demonstrated in Ref.~\cite{Beneke:2022rjv}. A coupling between $A$ and neutrinos would in principle open an additional annihilation channel, but such a process would only experience Sommerfeld enhancement and would remain subdominant compared to the super-resonant contribution we focus on here. Extensions of the benchmark model—for instance, by embedding it in a framework with an underlying gauge symmetry or in dark-meson scenarios where resonances naturally emerge in supersymmetric QCD~\cite{Csaki:2022xmu}—could provide UV completions.

\begin{figure}
    \centering
    \includegraphics[width=0.4\textwidth]{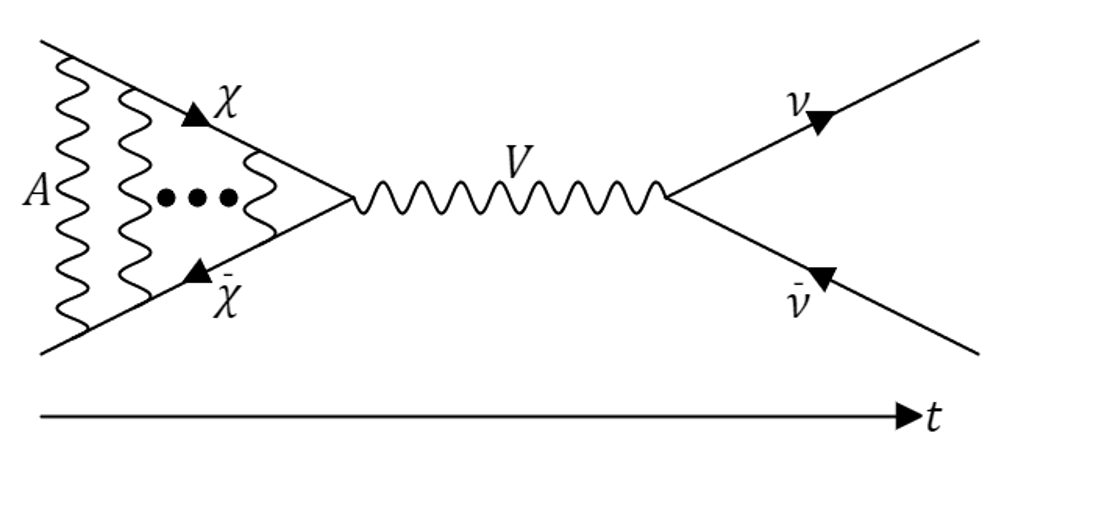}
    \caption{Leading Feynman diagram for DM super-resonant annihilation.}
    \label{fig:feynman}
\end{figure}

The possible DM annihilation channels in this model are $\bar{\chi} \chi \to A A$ and $\bar{\chi} \chi \to \bar{\nu} \nu$. The annihilation rate for $\bar{\chi} \chi \to A A$, when $v \ll 1$ and $m_A / m_\chi \ll 1$, is:
\begin{equation}
    \sigma v_{\rm rel} (\bar{\chi} \chi \to A A) \simeq \frac{\pi \alpha_A^2}{m_\chi^2} S_{\rm SF}(v), \label{eq:anni1}
\end{equation}
where $v_{\rm rel} = 2 v$, $\alpha_A = g_A^2 / 4\pi$, and $S_{\rm SF}(v)$ is the Sommerfeld enhancement factor for $s$-wave processes, mediated by $A$~\cite{0903.5307,0910.5713}:
\begin{equation}
    S_{\rm SF}(v) = \frac{\pi \alpha_A \sinh \left( \frac{12 m_\chi v}{\pi m_A} \right)}{v \left( \cosh \left( \frac{12 m_\chi v}{\pi m_A} \right) - \cos \left( 2 \pi \sqrt{\frac{6 m_\chi \alpha_A}{\pi^2 m_A} - \frac{36 m_\chi^2 v^2}{\pi^4 m_A^2}} \right) \right)}. \label{eq:sommerfeld}
\end{equation}
The annihilation rate for $\bar{\chi} \chi \to \bar{\nu} \nu$, the super-resonant process (see Fig.~\ref{fig:feynman}), is~\cite{Beneke:2022rjv}:
\begin{equation}
    \sigma v_{\rm rel} (\bar{\chi} \chi \to \bar{\nu} \nu) = \frac{g_V^2}{4 m_\chi^2} S_{\rm SF}(v) R(v), \label{eq:anniSRDM}
\end{equation}
with the dimensionless Breit-Wigner resonance factor:
\begin{equation}
    R(v) = \frac{m_\chi}{2} \frac{\Gamma_{\bar{\nu} \nu}}{(m_\chi v^2 - \delta M)^2 + \Gamma_V^2 / 4},
    \label{eq:resonance}
\end{equation}
where $\delta M = m_V - 2 m_\chi$, $\Gamma_V$ is the total decay width of $V$ (including $A A$, $\bar{\chi} \chi$, and $\bar{\nu} \nu$ channels), and $\Gamma_{\bar{\nu} \nu}$ is the decay width to the neutrino pair. Hereafter, we refer to the product $S_{\rm SF}(v) R(v)$ as the super-resonant enhancement factor (SREF) in our nomenclature. The decay width of $V$ to a fermion pair is:
\begin{equation}
    \Gamma_{\bar{f} f} = \frac{g^2}{12 \pi} m_V \left( 1 + \frac{2 m_f^2}{m_V^2} \right) \sqrt{1 - \frac{4 m_f^2}{m_V^2}}.
\end{equation}
Thus, $\Gamma_{\bar{\nu} \nu} \simeq \frac{g_\nu^2}{12 \pi} m_V$ when $m_\nu \ll m_V$, $\Gamma_{\bar{\chi} \chi} = \frac{g_V^2}{8 \pi} m_V \sqrt{\frac{2\delta M}{m_V}}$ if kinematically allowed, and $\Gamma_{A A}$ is loop-suppressed. For $\delta M \to 0$ and $g_V \sim g_\nu$, we approximate $\Gamma_V \approx \Gamma_{\bar{\nu} \nu}$.

While the relic density can be obtained from Eq.~\eqref{eq:nBE} using only the total $\sigma v_{\rm rel}$, solving the cBEs in Eq.~\eqref{eq:cBE} for greater accuracy requires the scattering amplitude $|\mathcal{M}|^2_{\chi \nu \leftrightarrow \chi \nu}$. In the non-relativistic limit, this is:
\begin{eqnarray}
    |\mathcal{M}|^2_{\chi \nu \leftrightarrow \chi \nu} &=& \frac{16 (g_\nu g_V)^2 (2 m_\chi / m_V)^2}{(2 m_\chi / m_V) t - 4 m_\chi^2} \beta (w, t) \simeq \frac{16 (g_\nu g_V)^2}{t - 4 m_\chi^2} \beta (w, t), \label{eq:ScattAmp}
\end{eqnarray}
where $\beta (w, t) = 8 m_\chi^2 w^2 + 4 m_\chi^2 (w / m_\chi + 1/2) t + t^2$, and the amplitude is summed over initial and final spin states, as well as particle and antiparticle states of $\nu$.

\section{Constraints and Prospects} \label{sec:constrain}

The Lagrangian in Eq.~\eqref{eq:lagrangian} has six free parameters. For simplicity, we impose three assumptions to reduce them to three: (1) $\delta = (2 m_\chi / m_V)^2 - 1 = 10^{-5}$ to enhance resonance; (2) $m_A = 6 \alpha_A m_\chi / (n^2 \pi)$ with $n = 2$ to boost Sommerfeld enhancement; (3) $\rho = g_V = g_\nu$. The remaining free parameters are $m_\chi$, $\rho$, and $\alpha_A$. 

\begin{figure}
    \centering
    \includegraphics[width=0.48\linewidth]{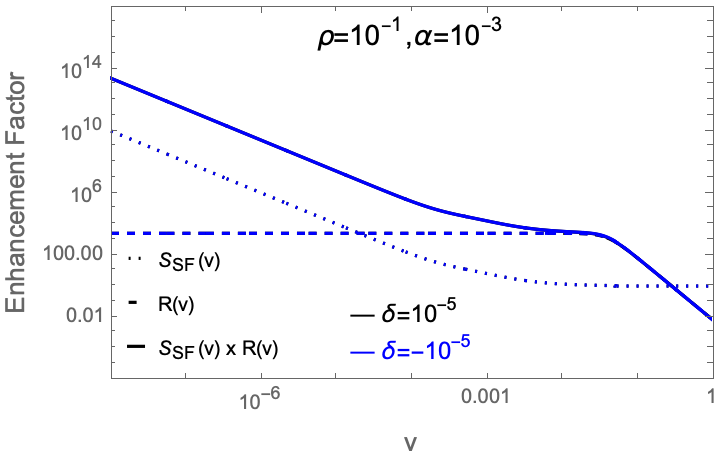}
    \includegraphics[width=0.48\linewidth]{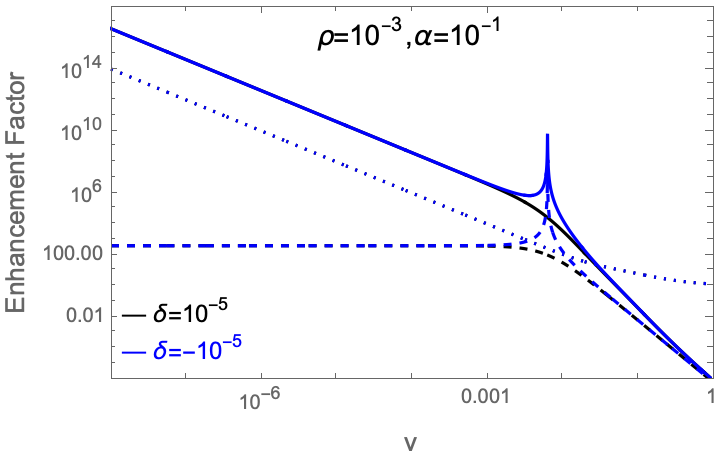}
    \includegraphics[width=0.48\linewidth]{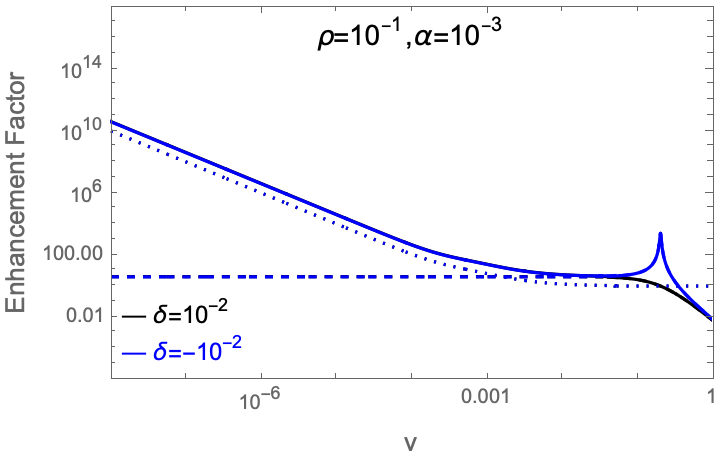}
    \includegraphics[width=0.48\linewidth]{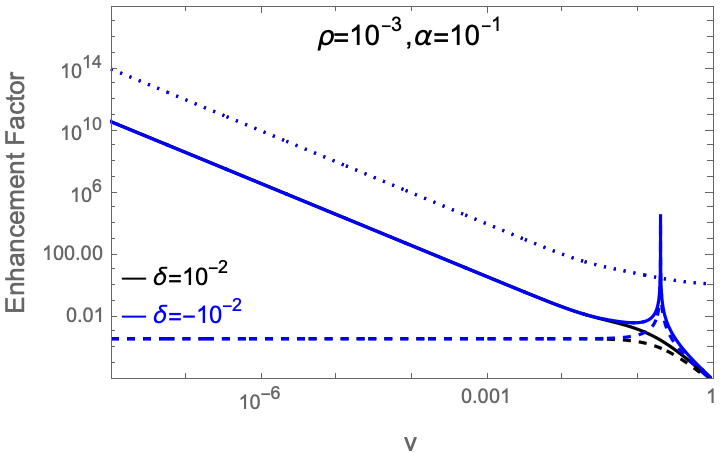}
    \caption{Sommerfeld ($S_{\rm SF}(v)$, dotted lines), resonance ($R(v)$, dashed lines), and SREFs ($S_{\rm SF}(v) \times R(v)$, solid lines) as a function of velocity for different $\rho$, $\alpha$, and $\delta$ values. DM mass is fixed to 1000 GeV. Black and blue lines are for positive and negative $\delta$ values respectively. Black lines are overlapped with blue lines if not appear in the sub-figures.}
    \label{fig:SR}
\end{figure}

In Fig.~\ref{fig:SR}, we present the Sommerfeld ($S_{\rm SF}(v)$), resonance ($R(v)$), and SREFs ($S_{\rm SF}(v) \times R(v)$) as a function of velocity for different $\rho$, $\alpha$, and $\delta$ values shown as dotted, dashed, and solid lines, respectively. DM mass is fixed to 1000 GeV. Black and blue lines are for positive and negative $\delta$ values respectively, Black lines are overlapped with blue lines if not appear in the sub-figures. At low DM velocities, the SREF is proportional to the inverse square of the velocity, $v^{-2}$, mirroring the behavior of the Sommerfeld enhancement factor. At higher velocities, the SREF scales as $v^{-4}$, consistent with the resonant factor in Eq.~\eqref{eq:resonance}, which also exhibits a $v^{-4}$ dependence when $\delta M$ and $\Gamma_V$ are sufficiently small. For smaller $|\delta|$ resonance factor become larger, and there is a narrow peak for negative $\delta$ compared with positive $\delta$.  The SREF is larger for larger $\rho$ or larger $\alpha$ when the other parameter is held fixed, since the resonant factor increases (for larger $\rho$) or the Sommerfeld enhancement becomes significant at higher velocities (for larger $\alpha$).

Both annihilation channels, $\bar{\chi} \chi \to A A$ (Eq.~\eqref{eq:anni1}) and $\bar{\chi} \chi \to \bar{\nu} \nu$ (Eq.~\eqref{eq:anniSRDM}), contribute to the relic density. However, since $\sigma v (\bar{\chi} \chi \to A A) / \sigma v (\bar{\chi} \chi \to \bar{\nu} \nu) \simeq \alpha_A^2 / 2$, we neglect the former process in the calculation.

With these assumptions, we use the super-resonant annihilation cross section (Eq.~\eqref{eq:anniSRDM}) and scattering amplitude (Eq.~\eqref{eq:ScattAmp}) to solve the nBE in Eq.~\eqref{eq:nBE} and cBEs in Eq.~\eqref{eq:cBE}. We have adapted the Mathematica package DRAKE~\cite{2103.01944} for precise and stable numerical computations.

Current and prospected constraints to the SRDM parameter space mainly arise from:
\begin{enumerate}
    \item \textbf{Big Bang Nucleosynthesis (BBN)}: In our SRDM model, the real vector $A$ decays via a DM loop to a neutrino pair, with a decay width of approximately $\Gamma_A \simeq 10^{-5} \alpha_A^2 \rho^4 m_\chi$. Requiring $\Gamma_A > 10^{-25}$ GeV ensures $A$ decays before Big Bang Nucleosynthesis (BBN)\footnote{Note that, if resonant particle is a real scalar $\phi$,  simple interaction lagrangian can be $\mathcal{L} \supset - g_A \bar \chi \gamma^\mu \chi A_\mu - g_ \phi\bar \chi  \chi \phi - g_\nu \bar \nu \nu \phi$. Then decay width of $A$ particle through DM one-loop to neutrino pair vanishes due to Ward identity, through DM triangle loop to two neutrino pairs are too small to decay before BBN.}~\cite{Cyburt:2002uv, Kawasaki:1994sc,Zhu:2021pad}.
    
    \item \textbf{Neutrino detection experiments}: Neutrino telescopes constrain dark matter models by searching for excess neutrinos produced in dark matter annihilations. If dark matter particles annihilate into neutrinos (either directly or via secondary channels), the resulting flux at Earth depends on the annihilation cross section and the underlying dark matter density profile. Experiments compare the predicted neutrino signal with observations, and in the absence of a significant excess above background, they set statistical upper bounds on the number of signal events, which are then translated into limits on the annihilation cross section. In this paper, we follow Ref.~\cite{Arguelles:2019ouk}, which provides updated constraints on the dark matter self-annihilation cross section into neutrinos using the most recent data, and forecasts the sensitivity of upcoming experiments. For consistency, the Navarro–Frenk–White (NFW) profile is adopted when presenting limits and projections.

    It is worth mentioning that electroweak corrections should be taken into account at high energy scales, even for the neutrino channel~\cite{Bauer:2020jay}. Such corrections not only slightly broaden the spectral peak above the electroweak scale but also generate a lower-energy continuum. While the former effect is not significant for high-energy neutrino detectors, the latter can enhance the flux at lower energies, potentially leading to stronger indirect detection bounds. However, the electroweak-corrected neutrino spectra differ only slightly from the neutrino-only results, implying that the impact on neutrino constraints remains modest~\cite{Liu:2020ckq}.
    
    \item \textbf{Cosmic neutrino background (C$\nu$B)}: While neutrinos produced from DM annihilation thermalize with the SM plasma before neutrino decoupling at $T \simeq 1\,\mathrm{MeV}$, any additional neutrino production after this epoch can affect the C$\nu$B, in particular modifying the effective number of neutrino species, $N_{\rm eff}$, and inducing spectral distortions of the C$\nu$B.
    
    $N_{\rm eff}$ quantifies the radiation energy density in the early universe relative to photons. Any injection of energy into neutrinos after their decoupling increases their temperature relative to the photon bath, leading to a positive contribution to $\Delta N_{\rm eff}$. This can be written as $\Delta N_{\rm eff} = \delta \rho_\nu / \rho_\nu$, where $\delta \rho_\nu$ is the additional energy density injected into neutrinos and $\rho_\nu$ is the standard neutrino energy density. In scenarios where dark matter annihilates exclusively into neutrinos after decoupling, the ratio $\delta \rho_\nu / \rho_\nu$ directly encodes the fractional energy injection, thereby translating into a measurable shift in $N_{\rm eff}$. In our SRDM model, this extra neutrino energy originates from dark matter depletion, i.e. $\delta \rho_\nu = \Delta \rho_\chi$, and $\Delta \rho_\chi$ can be obtained during the solving of the Boltzmann equations Eq.~\eqref{eq:nBE} and Eq.~\eqref{eq:cBE}. However, the value of $\Delta \rho_\chi$ obtained from cBEs is always larger than that obtained from the nBE  for the same parameter input, due to the additional dark matter annihilation accounted for in the cBEs--the relic density predicted by the cBEs is smaller than that from the nBE. Therefore, in our analysis, it is sufficient to calculate $\Delta \rho_\chi$ using the cBEs. We take the value of $\Delta N_{\rm eff} = 0.3$ and $\Delta N_{\rm eff} = 0.014$ for current and prospected sensitivities from Planck/DESI~\cite{Planck:2018vyg,DESI:2025ejh} and CMB-HD~\cite{CMB-HD:2022bsz} experiments, respectively.
    
    The spectral distortion of the C$\nu$B can be characterized in analogy to the photon case, with the distortion amplitude determined by the fractional energy injection into neutrinos. When additional energy is injected into the neutrino sector, such as from late dark matter annihilations, the relative increase in neutrino energy density, $\delta \rho_\nu / \rho_\nu$, quantifies the distortion of the otherwise thermal Fermi–Dirac spectrum. Determine $y$-parameter for neutrino through $y_\nu = \frac{1}{4} \frac{\delta \rho_\nu}{\rho_\nu}$, this parameter is just 1/4 of $\Delta N_{\rm eff}$. Given that the current limit is $y_\nu \leq 0.043$ \cite{Barenboim:2024wek}, we only have to consider limit of $y_\nu$ for constraining our model rather than of $\Delta N_{\rm eff} \leq 0.3$.

    \item \textbf{CMB}: Despite pure annihilation to neutrino pairs is essentially unconstrained by CMB anisotropy limits, because neutrinos free-stream and do not deposit energy into the gas that changes recombination. For high DM masses above electroweak breaking scale, neutrino final state is not strictly “invisible”: electroweak bremsstrahlung (emission of W/Z, photons, or charged leptons) produces electromagnetic particles which do deposit energy and can be constrained by the CMB. Especially for most of the parameter space in our model, annihilation cross section increases up to the CMB epoch. 
    
    The CMB upper bound on the annihilation cross section can be expressed as
    \begin{equation}
    \langle\sigma v\rangle \lesssim  m_\chi \frac{p_{\rm ann}}{f_{\rm eff}}, \label{eq:CMB}
    \end{equation}
    where $p_{\rm ann}$ is an effective parameter constrained by CMB anisotropies. We adopt $p_{\rm ann} = 3.2\times10^{-28},{\rm cm^3 s^{-1} GeV^{-1}}$ from the Planck 2018 results~\cite{Planck:2018vyg}. The efficiency factor $f_{\rm eff}$ was calculated with electroweak corrections in Ref.~\cite{Slatyer:2015jla}. Since the SM particle spectra from PPPC4DMID~\cite{Cirelli:2010xx} used in this reference are roughly a factor of two larger than recent detailed analyses of Ref.~\cite{Bauer:2020jay}, we conservatively divide $f_{\rm eff}$ by two (this factor does not significantly affect our results).

    For DM masses of $m_\chi = 100$, 1000, and 10000 GeV, the values of $f_{\rm eff}$ are 0, 0.01, and 0.05, respectively~\cite{Slatyer:2015jla}. Although $f_{\rm eff}$ for 100 TeV is not explicitly provided in the reference, it can be estimated to be at least 0.1, which we adopt for our analysis. Therefore, the corresponding upper limits on the annihilation cross section, obtained from Eq.~\eqref{eq:CMB}, are $+\infty$, $6.4 \times 10^{-20}$, $1.2 \times 10^{-20}$, and $6.4 \times 10^{-22} \, {\rm cm^3\,s^{-1}}$ for $m_\chi = 100$ GeV, 1 TeV, 10 TeV, and 100 TeV, respectively.

\end{enumerate}

\begin{figure}
    \centering
    \includegraphics[width=0.48\textwidth]{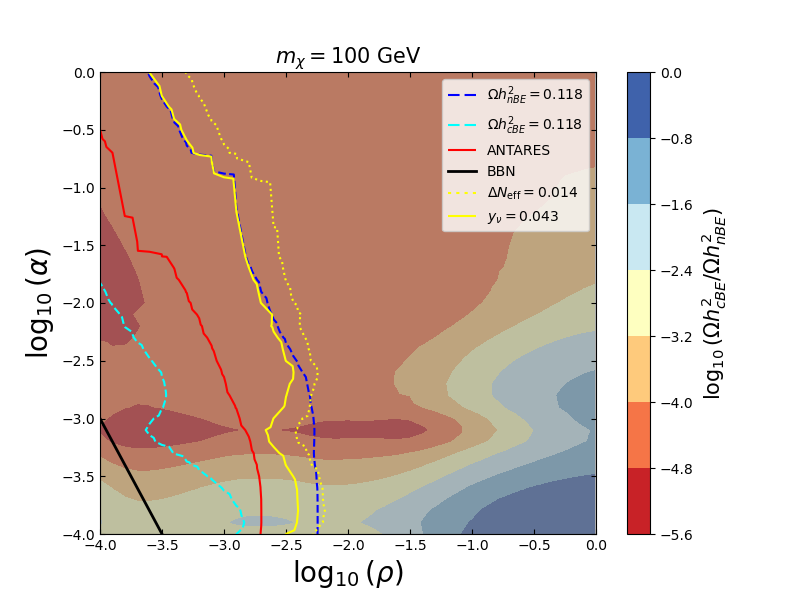}
    \includegraphics[width=0.48\textwidth]{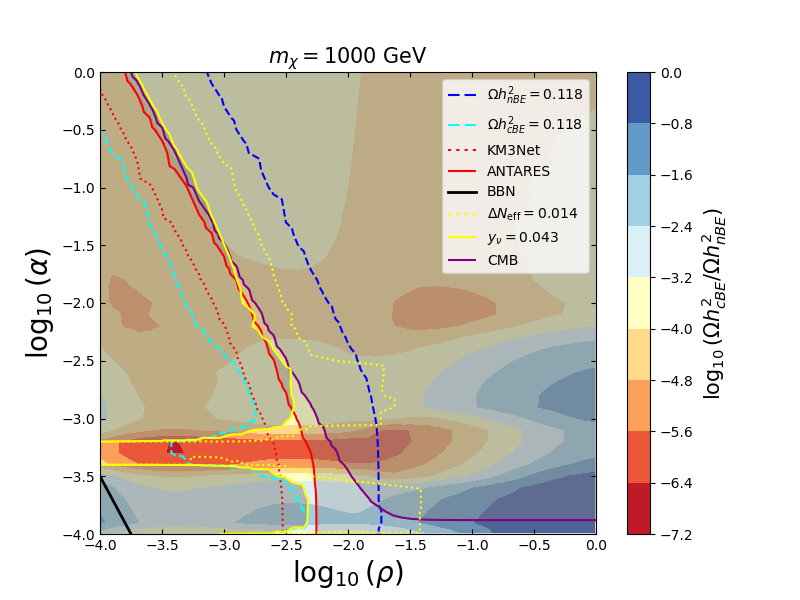}
    \includegraphics[width=0.48\textwidth]{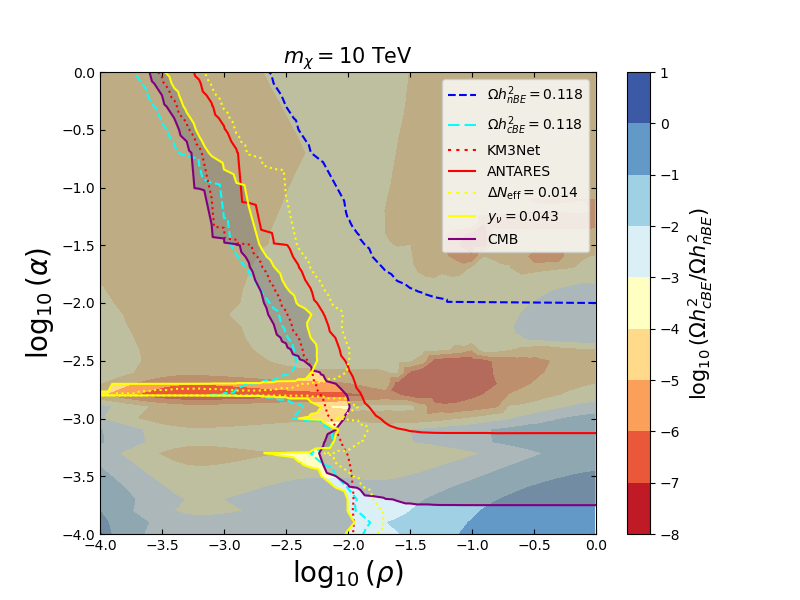}
    \includegraphics[width=0.48\textwidth]{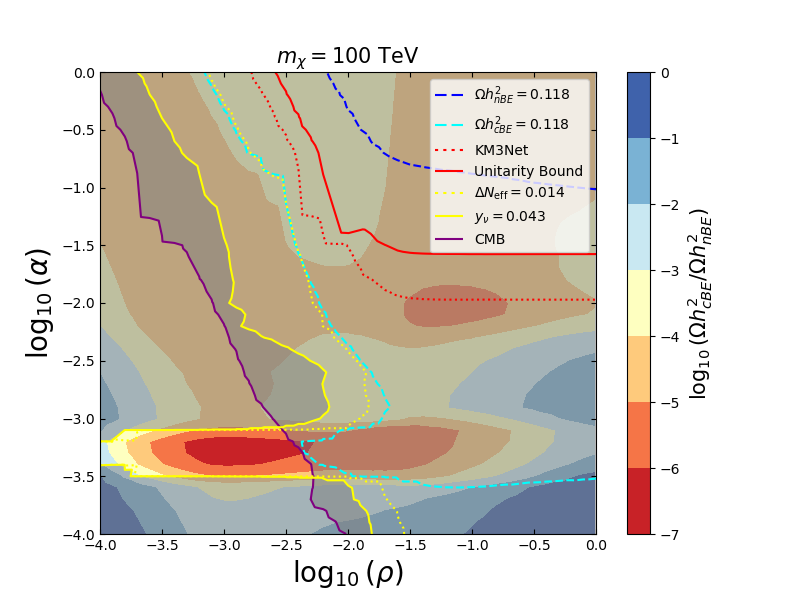}
    \caption{Viable and constrained parameter space in the $\rho$ vs. $\alpha_A$ plane for different DM masses. The color bar represents the ratio $\Omega h^2_{cBE} / \Omega h^2_{nBE}$, where $\Omega h^2_{cBE}$ and $\Omega h^2_{nBE}$ are relic densities from the cBEs and nBE, respectively. Dashed blue and green lines indicate $\Omega h^2_{cBE} = 0.118$ and $\Omega h^2_{nBE} = 0.118$, respectively. Solid red lines represent current upper limits for DM annihilation cross section from neutrino experiment ANTARES~\cite{Gozzini:2020dom} and "Unitarity Bound" for non-composite DM particle~\cite{Griest:1989wd,Smirnov:2019ngs}, while dotted red lines show future prospects from KM3Net~\cite{Gozzini:2020dom}. Thick black lines denote BBN constraints. Solid and dotted yellow line represent current limit and future sensitivities to the $y_\nu$ and $\Delta N_{\rm eff}$. Solid purple line represent limits from CMB. Currently excluded regions, right sides of solid purple/red lines and left sides of solid yellow lines, are shaded with gray. Annihilation cross section limits are: $\langle \sigma v \rangle_{\rm ANTARES} = 1.7 \times 10^{-24}$ cm$^3$ s$^{-1}$ for $m_\chi = 100$ GeV; $\langle \sigma v \rangle_{\rm ANTARES} = 1.1 \times 10^{-24}$ cm$^3$ s$^{-1}$, $\langle \sigma v \rangle_{\rm CMB} = 6.4 \times 10^{-20}$ cm$^3$ s$^{-1}$, and $\langle \sigma v \rangle_{\rm KM3Net} = 8.8 \times 10^{-26}$ cm$^3$ s$^{-1}$ for $m_\chi = 1000$ GeV; $\langle \sigma v \rangle_{\rm ANTARES} = 2.0 \times 10^{-24}$ cm$^3$ s$^{-1}$, $\langle \sigma v \rangle_{\rm CMB} = 1.2 \times 10^{-20}$ cm$^3$ s$^{-1}$, and $\langle \sigma v \rangle_{\rm KM3Net} = 1.4 \times 10^{-25}$ cm$^3$ s$^{-1}$ for $m_\chi = 10$ TeV; $\langle \sigma v \rangle_{\rm Unitarity\ bound} = 8.0 \times 10^{-24}$ cm$^3$ s$^{-1}$, $\langle \sigma v \rangle_{\rm CMB} = 6.4 \times 10^{-22}$ cm$^3$ s$^{-1}$, and $\langle \sigma v \rangle_{\rm KM3Net} = 1.4 \times 10^{-24}$ cm$^3$ s$^{-1}$ for $m_\chi = 100$ TeV.}
    \label{fig:result}
\end{figure}

Figure~\ref{fig:result} illustrates the viable parameter space in the $\rho$ vs. $\alpha_A$ plane for various DM masses. The color bar represents the relic density ratio $\Omega h^2_{cBE} / \Omega h^2_{nBE}$, derived from the cBEs and nBE, respectively. Dashed blue and green curves indicate where $\Omega h^2_{cBE} = 0.118$ and $\Omega h^2_{nBE} = 0.118$, matching Planck observations~\cite{Planck:2018vyg}. Solid and dotted red lines denote Unitarity bound/current 90\% confidence level limits and future sensitivities from neutrino experiment~\cite{Arguelles:2019ouk}. Note that, for DM masses above approximately 90 TeV, the Unitarity bound becomes more stringent than the current neutrino detection limits. Thick black lines represent BBN constraints from $A$ decay. Solid and dotted yellow lines represent current and future sensitivities to the $y_\nu-$distortion and $\Delta N_{\rm eff}$ from cosmological measurements respectively. Solid purple line represent limits from CMB.

For $m_\chi = 100$ GeV, the parameter space is fully excluded by ANTARES~\cite{ANTARES:2015vis,Albert:2016emp,Gozzini:2020dom} and $y_\nu$ constraints. For $m_\chi = 1000$ GeV, a small viable region exists between the constraints, and correct $\Omega h^2_{cBE}$ can be achieved within this region, suggesting a lower mass limit of $\mathcal{O}(100)$ GeV in our model if SRDM accounts for the full relic density within Planck’s $2\sigma$ range. While some of the parameter space of survived region can be reached by future KM3Net~\cite{Gozzini:2020dom} experiment, CMB-HD sensitivity fully cover the remaining region. The $m_\chi = 10$ TeV case is similar with $m_\chi = 1000$ GeV. For $m_\chi = 100$ TeV case, CMB limit rule out the DM relic density satisfying region. Therefore, SRDM mass in our model is constrained to the range of $\mathcal{O}(100)$ GeV–$\mathcal{O}(10)$ TeV if it constitutes 100\% of DM. This result indicates that the SRDM mass is constrained both from below and above: for smaller masses, extra energy injection into the neutrino background becomes prominent, while for higher masses, CMB constraints become stronger due to the increasing electroweak corrections. CMB observation put more stringent limits than neutrino indirect detection above $\sim 1$~TeV. In some regions of the parameter space, the relic density ratio $\Omega h^2_{cBE} / \Omega h^2_{nBE}$ can be as low as $\sim 10^{-7}$. Notably, as is clear from Fig.~\ref{fig:result}, using nBE instead of cBEs would lead to incorrect exclusion conclusions based on current experimental data.

Additionally, if DM annihilating to neutrinos constitutes only a fraction $f$ of the total DM, a larger annihilation cross section in the early universe is required~\cite{Zurek:2008qg, Herrero-Garcia:2018qnz}. In this scenario, the constraints from neutrino indirect detection experiments scale as $1/f^2$. Consequently, both the relic density curve and indirect detection limits shift toward larger cross sections, potentially allowing smaller DM masses.

\section{Conclusion} \label{sec:conclusion}
In this study, we investigated the SRDM model with a neutrino portal, focusing on its implications for relic density calculations and experimental constraints. Our primary objective was to assess how the super-resonant mechanism—comprising simultaneous enhancement of DM annihilation via narrow resonance and the Sommerfeld effect—affects relic density when kinetic and chemical decoupling occur concurrently. Using cBEs, we showed that the traditional nBE introduces significant inaccuracies in relic density predictions for this model, with discrepancies reaching $\Omega h^2_{cBE} / \Omega h^2_{nBE} \sim 10^{-7}$ in certain parameter regions.

Our analysis identifies a viable SRDM mass range of $\mathcal{O}(100)$ GeV to $\mathcal{O}(10)$ TeV, where it can account for all DM. Below this range, constraints from current neutrino detection experiments and spectral distortion $y_\nu-$parameter exclude the parameter space, while the CMB limits the upper end. Future experiments, such as KM3Net and CMB-HD, could probe portions of the remaining viable parameter space. Notably, reliance on the nBE rather than the cBEs would yield incorrect exclusion conclusions based on existing experimental data.

\section*{Acknowledgment}
This work is supported by the Natural Science Foundation of Xinjiang Uygur Autonomous Region of China (No. 2025D01C48), Tianchi Talent Project of Xinjiang Uygur Autonomous Region of China, and the National Natural Science Foundation of the People’s Republic of China (No. 12303002, 12275232).

\bibliography{refs}
\end{document}